\documentclass [11pt]{article}
\usepackage{amsmath}
\usepackage{amssymb}
\usepackage{latexsym}
\usepackage{color,url,framed}
\usepackage[color=yellow]{todonotes}
\usepackage{multicol}
\usepackage{graphicx}
\usepackage{amsfonts}
\usepackage[colorlinks]{hyperref}
\usepackage[margin=2cm]{geometry}

\newcommand{\ob}[1]{\overline{#1}}

\newcommand{\bm}[1]{\mbox{\boldmath{$#1$}}}
\newcommand{\bms}[1]{\footnotesize{\mbox{\boldmath{$#1$}}}}

\def\p{{\partial}}

\def\bx{{\bf x}}

\def\bR{{\bf R}}
\def\bu{{\bm{u}}}

\def\de{{\delta}}
\def\L{{\cal L}}

\def\ub{\ob{\bf u}}

\DeclareMathOperator{\diff}{d\!}

\newtheorem{theorem}{Theorem}[section]
\newtheorem{remark}[theorem]{Remark}

\title{
Stochastic Modelling in Fluid Dynamics: It\^o vs Stratonovich}
\author{
Darryl D. Holm\\
Imperial College London\\
email: d.holm@ic.ac.uk}
\date{
}

\begin{document}

\maketitle

\begin{abstract}
Suppose the observations of Lagrangian trajectories for fluid flow in some physical situation can be modelled sufficiently accurately by a spatially correlated It\^o stochastic process (with zero mean) obtained from data which is taken in fixed Eulerian space. Suppose we also want to apply Hamilton's principle to derive the stochastic fluid equations for this situation. Now, the variational calculus for applying Hamilton's principle requires the Stratonovich process, so we must transform from It\^o noise in the \emph{data frame} to the equivalent Stratonovich noise. However, the transformation from the It\^o process in the data frame to the corresponding Stratonovich process shifts the drift velocity of the transformed Lagrangian fluid trajectory out of the data frame into a non-inertial frame obtained from the It\^o correction. The issue is, ``Will non-inertial forces arising from this transformation of reference frames make a difference in the interpretation of the solution behaviour of the resulting stochastic equations?'' This issue will be resolved by elementary considerations.
\end{abstract}

\section{Introduction}


{\bf The Kelvin circulation theorem.}
The key element of fluid dynamics is the Kelvin circulation theorem, which is a statement of Newton's force law for mass distributed on closed material loops $c(\bm{u}_t^L)$ moving with the flow velocity vector, $\bm{u}_t^L(\bx)$, where the subscript $t$ denotes explicit time dependence. That is, the material loops move with the Lagrangian transport velocity vector $\bm{u}_t^L(\bx)$ tangent to the Lagrangian trajectory of the fluid parcel in the flow which occupies position $\bx$ at time $t$. 
For the fluid situation, Newton's law states that the time rate of change of the circulation integral -- around the material loop $c(\bm{u}_t^L)$ -- of the momentum-per-unit-mass co-vector $\bm{u^\flat}_t(\bx)$ is equal to the circulation of the co-vector representing force per unit mass $\bm{f^\flat}_t(\bx)$. Here, one denotes co-vector components by superscript $\flat$, as $\bm{u^\flat}_t(\bx)$. In integral form, this is
\begin{align}
\frac{d}{dt}\oint_{c(u_t^L)} u^\flat_t
=
\frac{d}{dt}\oint_{c(\bms{u}_t^L)} \bm{u^\flat}_t(\bx)\cdot d\bx
= \oint_{c(\bms{u}_t^L)} \bm{f^\flat}_t(\bx)\cdot d\bx 
=
\oint_{c(u_t^L)} f^\flat_t
\,,
\label{Kel-forcelaw}
\end{align}

where $u_t^L := \bm{u}_t^L(\bx)\cdot \nabla$ denotes the vector field with vector components $\bm{u}^L_t$  tangent to the Lagrangian trajectories in the vector field basis $\nabla$, and $u_t^\flat = \bm{u^\flat}_t\cdot d\bx$ denotes the circulation 1-form  with co-vector components $\bm{u^\flat}_t$ in the 1-form basis $d\bx$.

Physically, $\bm{u}_t(\bx)$ is the time-dependent \emph{momentum per unit mass} measured in a fixed Eulerian frame. Since momentum per unit mass and velocity have the same dimensions and because momentum and force are defined in Newton's force law to be measured in an inertial reference frame, one may refer to $\bm{u}_t(\bx)$ (without superscript $\flat$) as the \emph{Eulerian velocity}.  
Thus, the Kelvin-Newton relation in \eqref{Kel-forcelaw} for loop momentum dynamics involves two kinds of `velocity', both of which may be evaluated at a given point $\bx\in\mathbb{R}^3$ in an inertial frame with fixed Eulerian coordinates. However, the vector field $u_t^L$ and the 1-form $u^\flat_t$  in Kelvin's theorem have quite different transformation properties. 


As we said, the vector field $\bm{u}_t^L(\bx)$ is the velocity at each point $\bx$ fixed in space \emph{along the path} of the material masses distributed in the line elements along the moving loop. Thus, the velocity $\bm{u}_t^L(\bx)$ in \eqref{Kel-forcelaw} may be regarded as a Lagrangian quantity, because its argument is the pullback of the tangent to the Lagrangian trajectories of the fluid parcels of the circulation loop moving through fixed Eulerian space under the smooth invertible flow map, $\bx_t=\phi_t\bx_0$, where $\phi_0=Id$. That is, 
\begin{align}
\frac{d}{dt}\phi_t(\bx_0) = \phi_t^*{u}^L(t,\bx_0) := {u}^L(t,\phi_t (\bx_0))={u}^L_t(\bx_t)
\,.
\label{uL-def}
\end{align}
Here, $u_t^L := \bm{u}_t^L(\bx)\cdot \nabla$ is the vector field tangent to the Lagrangian trajectories. In contrast, the circulation 1-form $u_t^\flat = \bm{u^\flat}_t\cdot d\bx$ has co-vector components -- denoted by superscript $\flat$ -- as $\bm{u^\flat}_t(\bx)$ and representing the momentum per unit mass at position $\bx$ at time $t$ which is determined from Newton's laws of motion.

We stress that the Kelvin-Newton relation \eqref{Kel-forcelaw} is a statement about the time rate of change of the momentum-per-unit-mass 1-form $u_t^\flat$ distributed on closed material loops. In particular, Kelvin circulation is not about the \emph{acceleration} of velocity distributions $u_t^L$ on closed loops. Unfortunately, this distinction can often be lost for fluid motion in an inertial frame, because the momentum is simply proportional to the velocity in that case, and in the $\mathbb{R}^3$ inner product the distinction between  vector fields $u_t^L$ and 1-forms $u_t$ is a nicety, \emph{except for their transformations under smooth maps}. Therefore, in transforming to a noninertial frame such as a rotating frame, the distinction becomes important even in $\mathbb{R}^3$. In that case, the velocity $u_t^L := \bm{u}_t^L(\bx)\cdot \nabla$ is the velocity vector field \emph{relative} to the reference frame moving at velocity $\bR(\bx)$, while $u_t = (u^L_t)^\flat(\bx) + \bm{R^\flat}(\bx)\cdot d\bx$ is the \emph{total} momentum per unit mass, relative to the fixed frame after being parallel transported to the coordinate system $\bx$ in the moving frame by using the \emph{connection 1-form} $\bm{R^\flat}(\bx)\cdot d\bx$.

Thus, the Coriolis force arises in the acceleration of the \emph{relative velocity} in the moving reference frame with coordinates $\bx$ fixed on the surface of the rotating Earth. The Coriolis parameter is ${\rm curl}\bm{R}(\bx)= 2 \bm{\Omega}(\bx)$ where $ \bm{\Omega}(\bx)$ is the angular velocity of the Earth in the moving frame, relative to the fixed stars. Newton's force law for the rate of change of total momentum in \eqref{Kel-forcelaw} becomes
\begin{align}
 \frac{d}{dt}\oint_{c(\bms{u}_t^L)} \big(\bm{u}_t^L(\bx)+ \bm{R}(\bx)\big)\bm{^\flat}\cdot d\bx
= \oint_{c(\bms{u}_t^L)} \bm{f^\flat}_t(\bx)\cdot d\bx 
\,,
\label{Kel-forcelaw-conc1}
\end{align}
while Newton's relation between acceleration of the relative velocity and force becomes
\begin{align}
 \frac{d}{dt}\oint_{c({\bms{u}}_t^L)} (\bm{u}_t^L)\bm{^\flat}(\bx)\cdot d\bx
= \oint_{c(\bms{u}_t^L)} \big(\bm{f}_t(\bx) + \bm{u}_t^L \times 2 \bm{\Omega}\big)\bm{^\flat}\cdot d\bx 
\,.
\label{Kel-accelaw-conc1}
\end{align}
Thus, the circulation of Newton's force law in terms of momentum in \eqref{Kel-forcelaw-conc1} is covariant under the  change of reference frame, while the circulation of Newton's force law in terms of acceleration in \eqref{Kel-accelaw-conc1} changes its form by acquiring the `fictitious' Coriolis force. That is, the form of Newton's law  $F=ma$ is covariant under changes of frame, only if $ma=dP/dt$. Having made this point in the context of Kelvin's circulation theorem, we will henceforth drop the superscript $\flat$ for 1-forms and assume that the reader will understand the differences between vector fields and 1-forms in context in the remainder of the paper. 

{\bf Background of the stochastic problem.}
The form of the Kelvin circulation theorem in \eqref{Kel-forcelaw-conc1} persists for stochastic flow, provided the Lagrangian paths follow Stratonovich stochastic paths, as shown in \cite{Holm2015} by using a Stratonovich stochastic version of Hamilton's principle for fluid dynamics  in an inertial domain. The observation of the persistence of the Kelvin form \eqref{Kel-forcelaw-conc1} for Stratonovich stochastic fluid trajectories has led to the SALT algorithm for uncertainty quantification and data assimilation for stochastic fluid models.%
\footnote{SALT is an acronym for Stochastic Advection by Lie Transport \cite{SGFDclub2019a,SGFDclub2019b}.} 
The SALT algorithm proceeds from data acquisition, to coarse graining, to uncertainty quantification by using stochastic fluid dynamical modelling. The algorithm then continues to uncertainty reduction via data assimilation based on particle filtering methods, as discussed and applied in  \cite{SGFDclub2019a,SGFDclub2019b}. 

\begin{figure}[h!]
    \centering
    {\includegraphics[scale=0.5]{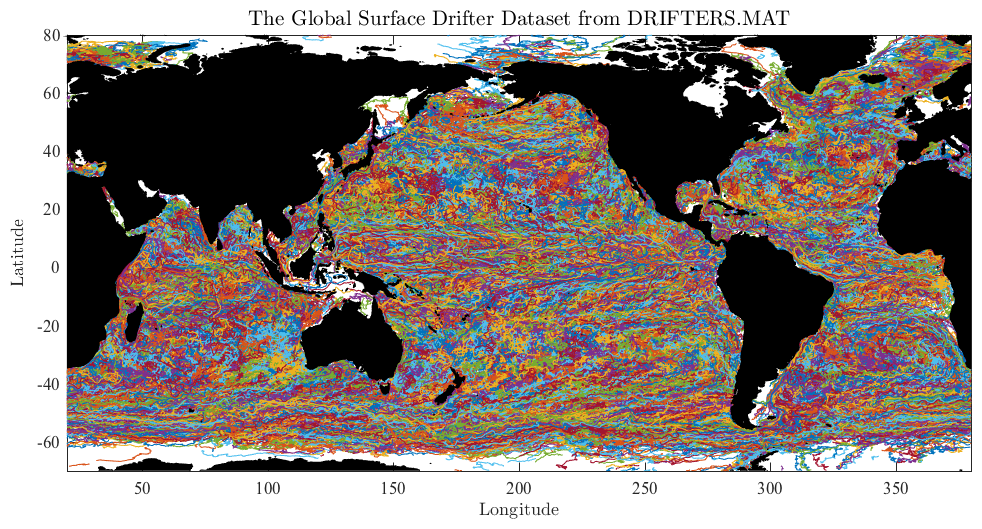} \label{fig:globaldrift}}
    \caption{Trajectories from the National Oceanic and Atmospheric Administration Global Drifter Program are shown, in which each colour corresponds to a different drifter.}
\end{figure}

The present note has a simple storyline. Suppose the Lagrangian trajectories for fluid flow in some physical situation are modelled sufficiently accurately by a spatially correlated It\^o stochastic process obtained from data which is taken to be statistically stationary with zero mean in the inertial frame of fixed Eulerian space. For example, this could be drifter data on the surface of the Ocean as seen from a satellite, as shown in Figure \ref{fig:globaldrift}.%
\footnote{The spatial correlations of the data shown in Figure \ref{fig:globaldrift} depend on the season, which can be modelled as a prescribed long-term time-dependence. However, we neglect such prescribed time-dependence here, to simplify the presentation.}

\begin{remark}
Figure \ref{fig:globaldrift} \cite{Lilly2017}  
displays the global array of surface drifter trajectories from the National Oceanic and Atmospheric Administration's ``Global Drifter Program'' (\url{www.aoml.noaa.gov/phod/dac}). In total, more than 10,000 drifters have been deployed since 1979, representing nearly 30 million data points of positions along the Lagrangian paths of the drifters at six-hour intervals. An important feature of this data is that the ocean currents show up as spatial correlations, easily recognised visually by the concentrations of colours representing individual paths in Figure \ref{fig:globaldrift}. These spatial correlations exhibit a variety of spatial scales for the trajectories of the drifters, corresponding to the variety of spatiotemporal scales in the evolution of the ocean currents which transport the drifters. \hfill $\Box$
\end{remark}

Suppose we \emph{also} want to apply Hamilton's principle to derive the stochastic fluid equations for this situation. Now, the variational calculus for applying Hamilton's principle requires the chain rule and the product rule from vector calculus. The Stratonovich process respects these calculus rules, but It\^o calculus is another matter. Thus, to use these rules of calculus to apply Hamilton's principle, we should transform from It\^o noise with zero mean in the \emph{data frame} to the equivalent Stratonovich noise. 

{\bf Problem statement.}
The question is, `Will the transformation transform from It\^o noise in the \emph{data frame} to the equivalent Stratonovich noise make a difference in the solution behaviour of the resulting stochastic equations?' 

{\bf Framework for resolving this issue.}
We know the transformation from the It\^o process in the data frame to the corresponding Stratonovich process shifts the drift velocity of the transformed Lagrangian fluid trajectory out of the data frame into a non-inertial frame obtained from the It\^o correction. We know the It\^o correction explicitly, since the spatial correlations of the It\^o noise have been obtained from the observed data. So, perhaps all is well, even though the spatial correlations depend upon location. 

Thus, we have seen that the It\^o correction shifts the Stratonovich drift velocity of the fluid into a spatially-dependent non-inertial frame relative to the \emph{data frame}. (The data frame is the fixed Eulerian frame in which the It\^o drift velocity was defined.) Now, such a shift of reference frame would introduce a non-inertial force into the motion equation for $\bm{u}_t$, whose derivation using variational calculus requires the Stratonovich representation of the noise. According to oceanographic experience, this non-inertial force can generate circulation of the Eulerian velocity \cite{CL1976,C85}. The question then arises, `Is the 3D circulation which would be generated by the non-inertial force due to the It\^o correction going to be important the the comparison of the stochastic motion equation to the observed fluid motion?' 

To demonstrate the resolution of this issue, we apply Hamilton's principle to derive the equations of motion in the example of the stochastic Euler-Boussinesq equations (SEB) for the incompressible flow of a stratified fluid under gravity.%
\footnote{In Hamilton's principle, such shifts of frame are accomplished transparently by an additive term in the action integral.}
 In this case, including the non-inertial force produces a ``vortex force" analogous to the Coriolis force. Upon inspection, we will recognise the derived equations as a version of the Craik-Leibovich equations \cite{C85,CL1976}, altered by the presence of stochastic advection by Lie transport (SALT). 

{\bf Oceanographic background.} The ``vortex force" of the deterministic Craik-Leibovich (DCL) theory derived in \cite{CL1976,C85} was introduced to model the observed phenomenon of Langmuir circulations arising physically from wave--current interaction (WCI), \cite{L1}-\cite{LT}. The corresponding velocity shift due to WCI was called the ``Stokes mean drift velocity" and was a \emph{prescribed quantity} denoted as $\ub^S(\bx)$. The importance of including $\ub^S(\bx)$ in the DCL equations has been investigated for Kelvin-Helmholtz instability in \cite{Holm1996} and for symmetric and geostrophic instabilities in the wave-forced ocean mixed layer in \cite{HaneyF-K2015}. In fact, because of its effectiveness in generating Langmuir circulations, the DCL has become a standard feature of the wave--current interaction (WCI) literature. 

The three-dimensional  results of having transformed the stochastic Euler-Boussinesq (SEB) fluid equations into a stochastic version of Craik-Leibovich equations have yet to be investigated. 
However, it would not be surprising if the solutions of the Craik-Leibovich (SCL) equations were interpreted as possessing Langmuir circulations generated by the It\^o correction to the stochastic drift velocity. Such an interpretation should be received with care, though, since they would represent circulations of the \emph{relative velocity}, $\bm{u}^L$, generated simply because the equations for $\bm{u}^L$ are not written in the inertial frame of the data. 

{\bf Objective of the paper.}
The present note investigates how to deal with non-inertial forces in stochastic dynamics which arise from It\^o corrections as changes of frame when applying mixed It\^o and Stratonovich stochastic modelling in 3D stochastic Euler-Boussinesq (SEB) fluid dynamics. 

The resolution of this issue has already been given above in the comparison between equations \eqref{Kel-forcelaw-conc1} and \eqref{Kel-accelaw-conc1}.  Namely, the It\^o correction will generate no Langmuir circulations, as seen in the \emph{data frame} with `velocity' $\bm{u}_t(\bx)=\bm{u}_t^L(\bx)-\ub^S(\bx)$, which is really the momentum per unit mass. However, Langmuir circulations would indeed be viewed in the relative drift frame of the \emph{Lagrangian} fluid parcels with velocity $\bm{u}_t^L(\bx)$, as being caused by the non-inertial force felt in the moving frame of the It\^o correction $\ub^S(\bx)$. The presence of this sort of fictitious force is why Newton's law of motion $F=ma$ only applies in an inertial frame. Undergraduate physics students will recognise this point as the analogue of the familiar Coriolis force felt in a rotating frame. They may also recall that the canonical momentum is not necessarily the mass times the velocity in a rotating frame, or in an external magnetic field. Although the reasoning in the remainder of the paper is elementary, we hope the explicit stochastic fluid dynamical calculations which  demonstrate the resolution of this issue for 3D stochastic Euler-Boussinesq (SEB) fluid dynamics below may be illuminating. 


\subsection{Stochastic Kelvin circulation dynamics}

This section describes the background for the It\^o correction in stochastic fluid dynamics. 

Multi-time homogenisation for fluid dynamics in \cite{CGH2017} was used to derive the following It\^o representation of the stochastic vector field which generates a stochastic Lagrangian fluid trajectory in the Eulerian representation,
\begin{align}
\diff \bx_t = \bm{u}_t(\bx_t)\,dt + \bm{\xi}(\bx_t)\,dB_t
\,,\quad \hbox{with}\quad {\rm div}\bm{\xi} = 0
\,,
\label{ito-traj}
\end{align}
where subscript $t$ denotes explicit time dependence, i.e., not partial time derivative.  In this notation, $dB_t$ denotes a Brownian motion in time, $t$, whose divergence-free vector amplitude $\bm{\xi}(\bx_t)$ depends on the Eulerian spatial position $\bx_t\in \mathbb{R}^3$ along the Lagrangian trajectory, $\bx_t=\phi_t\bx_0$ with initial condition $\bx_0$.
The differential notation ($\diff\,$) in equation \eqref{ito-traj} is short for 
\begin{align}
\bx_t  - \bx_0 = \int_0^t \diff \bx_t 
= \int_0^t  \bm{u}_t(\bx_t)\,dt + \int_0^t \bm{\xi}(\bx_t)\,dB_t
\,,\label{integrated-traj}
\end{align}
where the first time integral in the sum on the right is a Lebesque integral and the second one is an It\^o integral. 
The representation of stochastic Lagrangian fluid trajectories in equation \eqref{ito-traj} has a long history, going back at least to GI Taylor \cite{Taylor1921}, who provided an exact Lagrangian solution for the rate of spread of tracer concentration in unbounded, stationary homogeneous turbulence. Equation \eqref{ito-traj} is also a fundamental tenet in atmospheric science. See \cite{ThomsonWilson2013} for a historical survey of the applications of this ansatz in atmospheric science. Let us also mention a few recent papers which are more directly related to our present lines of thought about fluid dynamics with multiplicative noise  \cite{Brzeniak1992,Falkovich2001,Flandoli2011,Memin2014,MikRoz2004,MikRoz2005}.

The Stratonovich representation  (denoted with symbol $\circ$) of the It\^o trajectory in \eqref{ito-traj} is given by transforming to
\begin{align}
\diff \bx^L_t = \bu^L_t(\bx_t)\,dt + \bm{\xi}(\bx_t)\circ\,dB_t
\,,\quad \hbox{with}\quad {\rm div}\bm{\xi} = 0
\,.
\label{strat-traj}
\end{align}

\begin{remark}\label{rmd-strat-int}
The quantity ${\rm d} x_t(x)$ in \eqref{strat-traj} may be regarded as a stochastic Eulerian vector field which generates a smooth invertible map in space whose parameterisation in time is stochastic. In integral form, the operation the expression ${\rm d} \bx_t$ in \eqref{strat-traj} represents,%
\footnote{The usual superscript $\omega$ for pathwise stochastic quantities will be understood throughout. However, this superscript will be suppressed for the sake of cleaner notation.} 
\begin{equation}
\bx_t = \bx_0 + \int_0^t \bu(\bx,t)\,dt +  \int_0^t \boldsymbol{\xi}(\bx)\circ dB(t)\,,
\label{d-def-strat-integral}
\end{equation}
which is the sum of a Lebesgue integral and a Stratonovich stochastic integral.
\hfill ${\Box}$
\end{remark}

The difference in drift velocities for the two equivalent representations \eqref{ito-traj} and \eqref{strat-traj} of the same Lagrangian trajectory $\diff \bx^L_t = \diff \bx_t $ is called the It\^o correction \cite{Gardiner}. It is given by,
\begin{align}
\bu^L_t(\bx_t) - \bu_t(\bx_t) 
= - \frac12 \big(\bm{\xi}(\bx_t)\cdot\nabla\big) \bm{\xi}(\bx_t) 
=: \bu^S(\bx_t)\,.
\label{StokesDrift-ansatz}
\end{align}
The difference of velocities $\bu^S=\bu^L_t - \bu_t$ is called the ``It\^o-Stokes drift velocity'' in \cite{Bauer-etal2019}, as an analogue of the classic Stokes mean drift velocity, which is traditionally written as $\ub^S(\bx)=\ub^L(\bx,t) - \ub^E(\bx,t)$, where the overlines denote time averages or phase averages at constant Lagrangian and Eulerian positions, respectively. Identifying the difference $\bu^S=\bu^L_t - \bu_t$ in equation \eqref{StokesDrift-ansatz} with the traditional Stokes mean drift velocity $\ub^S$ in the Deterministic Craik-Leibovich (DCL) model emphasises the potential physical importance of the choice between It\^o and Stratonovich noise in modelling uncertainty in fluid dynamics. Note, however, that $\bu^S(\bx)$ (without overline) is the It\^o correction, while $\ub^S(\bx)$ (with overline) is the Stokes mean drift velocity. Although the notation stresses the analogue, the distinction between $\bu^S(\bx)$ in \eqref{StokesDrift-ansatz} and $\ub^S(\bx)$ should be clear in context. 

\begin{remark}[Physical implications of the Stokes mean drift velocity]\label{Stokes+CL-remark}$\,$\\
The traditional Stokes mean drift velocity $\ub^S(\bx)$ is assumed to be a time-independent prescribed difference between the Lagrangian mean fluid velocity $\ub^L(\bx,t)$ and its Eulerian mean counterpart $\ub^E(\bx,t)$ \cite{Stokes1847}. The Stokes drift velocity plays a key role in the celebrated Craik-Leibovich (CL) model of Langmuir circulations arising from wind forcing at the air-sea interface in oceanography \cite{CL1976}. See \cite{Thorpe2004} for a review of recent advances in modelling and observing Langmuir circulations driven by wind and waves in the upper layers of large bodies of water.  

In the present notation, $\bu_t$ is the Eulerian momentum per unit mass in Newton's 2nd law and  $\bu^L_t$ is the transport drift velocity for the corresponding equivalent Stratonovich representation of the Lagrangian trajectory. The difference $\bu^S(\bx)$ between these two quantities in \eqref{StokesDrift-ansatz} with the dimension of velocity at a fixed Eulerian point $\bx$ along the Lagrangian trajectory $\bx_t=\phi_t\bx_0$ may be assumed to be time-independent, provided the statistics of the observed data is stationary. One may also prescribe a temporal dependence of the It\^o correction $\bu^S(\bx)$ to vary with the seasons in geophysical applications, say, without interfering with the conclusions of the SCL model. 

The effects of uncertainty in the statistics of the Stokes mean drift velocity $\ub^S$ in the context of the Craik-Leibovich model has been treated in \cite{Holm2020}, as well. However, no self consistent dynamical theory of the Stokes drift $\ub^S$ has been developed yet, to our knowledge. Nonetheless. today, the Stokes drift representation of the wave-current interaction (WCI) in the Euler-Boussinesq (EB) fluid motion equation is in general use for numerically modelling the vertical transport effects of Langmuir circulations on mixed layer turbulence by using large-eddy simulations (LES) approach in computational fluid dynamics. However, the theoretical issues are by no means settled. For a recent discussion of these issues from the viewpoint of LES computations, see, e.g., \cite{Fujiwara-etal2018,Fujiwara-MellorReply2019,Mellor-Fujiwara2019,Tejada-Martinez2020}. \hfill $\Box$
\end{remark}

The Kelvin circulation integral for the Eulerian representation of the Lagrangian trajectory in \eqref{ito-traj} is defined as
\begin{align}
I(t) = \oint_{c(\diff \bx^L_t)} \bm{u}_t\cdot d\bx
\,,\label{KelCirc-Ito}
\end{align}
where $\bm{u}_t(\bx)$ is the Eulerian velocity at a fixed spatial position $\bx\in \mathbb{R}^3$ and $\diff \bx^L_t $  is the Stratonovich representation of the transport velocity of the circulation loop moving along the Lagrangian trajectory determined by integrating the semimartingale relationship in the vector field \eqref{ito-traj} to find the path \eqref{integrated-traj}. 

In the Stratonovich representation of the transport velocity vector field $\diff \bx^L_t$ in \eqref{strat-traj}, we may use the ordinary rules of calculus to compute the evolution equation for the circulation in equation \eqref{KelCirc-Ito}.
For this calculation, we invoke the evolutionary version of the classic Kunita-It\^o-Wentzell (KIW) formula \cite{Ku1981,Ku1984,Ku1997} for a 1-form, as derived in \cite{BdlHLT2019}. The KIW formula produces the following dynamics,
\begin{align}
\begin{split}
\diff \oint_{c(\diff \bx_t^L)} \bm{u}_t\cdot d\bx
&=
\oint_{c(\diff \bx_t^L)} \big(\diff + \L_{\diff \bx_t^L}  \big) (\bm{u}_t\cdot d\bx)
\\&=
\oint_{c(\diff \bx_t^L)} \big(\diff \bm{u}_t + (\diff \bx_t^L\cdot\nabla ) \bm{u}_t  
+ (\nabla \diff \bx_t^L)^T \cdot \bm{u}_t) \big) \cdot d\bx
\\&=
\oint_{c(\diff \bx_t^L)} \big(\diff \bm{u}_t - \diff \bx_t^L\times {\rm curl}\bm{u}_t  
+ \nabla (\diff \bx_t^L\cdot \bm{u}_t) \big) \cdot d\bx
\,,
\end{split}
\label{KelCirc-Strat}
\end{align}
where the operator $ \L_{\diff \bx_t^L} $ denotes the \emph{Lie derivative} with respect to the vector field $\diff \bx_t^L$. 
Equation \eqref{KelCirc-Strat} will play a role in deriving the Kelvin circulation theorem, itself, and thereby interpreting the solution behaviour of the fluid motion equation, derived below from Hamilton's principle. 

In the next section, we will show how passing from the It\^o representation of the Lagrangian trajectory in \eqref{ito-traj} to its equivalent Stratonovich representation in \eqref{strat-traj} enables the use of variational calculus to derive the equations of stochastic  fluid motion via the approach of stochastic advection by Lie transport (SALT), based on Hamilton's variational principle using Stratonovich calculus, \cite{Holm2015}. The resulting equations will raise the issue of non-inertial forces and this issue will be resolved by elementary considerations.

\section{SALT derivation of stochastic Euler-Boussinesq (SEB)}\label{appB}

\subsection{Hamilton's principle, motion equations and circulation theorems for SALT}
Following \cite{Holm2015} we apply Hamilton's principle $\delta S = 0$ with the following action integral $S=\int_0^T \ell(\bu_t^L,D,b) \,dt$ whose fluid Lagrangian $\ell(\bu_t^L,D,b)$ depending on drift velocity $\bu_t^L$, buoyancy function $b(\bx,t)$ and the density $D(\bx,t)d^3x$ for $(\bx,t)\in \mathbb{R}^3\times\mathbb{R}$. We constrain the variations to respect the \emph{stochastic} advection equations with transport velocity $\diff \bx^L_t$ given in \eqref{strat-traj},
\begin{align}
 {\diff\,} b + \diff \bx^L_t \cdot \nabla b = 0
\,,\quad \hbox{and} \quad
{\diff\,} D + {\rm div}(D\diff \bx^L_t) = 0
\,.\label{aux-stoch}
\end{align}
These relations ensure that the values of the advected quantities $b$ and $D(\bx,t)d^3x$ remain invariant along flow given by the stochastic Lagrangian trajectory in \eqref{integrated-traj}. 

In general, with the constraints in \eqref{aux-stoch} Hamilton's principle will result in a motion equation in the Euler-Poincar\'e form \cite{HMR1998}
\begin{align}
\big(\diff + \L_{\diff \bx_t^L}  \big)\big( \bu_t \cdot d\bx \big)
=  \frac1D \frac{\de \ell}{\de b} db + d \frac{\de \ell}{\de D}
\quad\hbox{with}\quad 
\bu_t := \frac1D \frac{\de \ell}{\de \bu_t^L}
\,.\label{EP-stoch}
\end{align}
The various differential operators in \eqref{EP-stoch} are defined, as follows. As usual, $d$ denotes spatial differential
of functions, e.g., $db=\nabla b\cdot b\bx$. Likewise, $\delta$ denotes variational (Gateux) derivative of functionals,
e.g., $\delta \ell(u) = \langle \delta \ell/\delta u\,,\, \delta u \rangle$ where $\langle\,\cdot\,,\,\cdot\,\rangle$ denotes $L^2$ 
pairing. Finally, Roman $\rm d$ denotes stochastic `differential', in the sense of stochastic integrals defined in remark \ref{rmd-strat-int}, cf. formulas \eqref{strat-traj} and \eqref{d-def-strat-integral}.

The stochastic Euler-Poincar\'e equation \eqref{EP-stoch} will result in a Kelvin-Newton theorem of the form 
\begin{align}
\begin{split}
\diff \oint_{c(\diff \bx_t^L)} \!\!\!\big( \bu_t \cdot d\bx \big)
&=
\oint_{c(\diff \bx_t^L)} 
 \frac1D \frac{\de \ell}{\de b} db 
 + \oint_{c(\diff \bx_t^L)}  d \frac{\de \ell}{\de D}
\end{split}
\,,\label{KelCirc-Strat-EP}
\end{align}
and the loop integral of an exact differential in the last termwill vanish. 
For more discussion of stochastic advection, see \cite{BdlHLT2019}.
For discussion of other stochastic Kelvin theorems, see \cite{DrivasHolm2019}.

For the example of the stochastic Euler-Boussinesq (SEB) equations, the
pressure constraint appearing in the well known deterministic action integral \cite{Holm1996}
must be altered to become, 
\begin{align}
S=\int_0^T \ell(\bu^L_t,D,b) \,dt =  \int dt \int d^3 x \left[ \frac12 D |\bu^L_t|^2 - D \bu^L_t \cdot \bu^S(\bx)
 - gDbz\right] -  \int d^3 x  \int {\rm d}p(D-1),
\label{lag-stoch}
\end{align}
and again constrain the variations by requiring satisfaction of the stochastic advection relations in \eqref{aux-stoch}. Special care is required when imposing the incompressibility constraint, ${\rm div}(\diff \bx^L_t)=0$  by requiring that $(D=1)$, since the quantity $D$ is a stochastic quantity. As explained in  \cite{SC2020}, this means the pressure Lagrange multiplier $({\rm d}p)$ is a semimartingale.%
\footnote{The validity of the incompressibility relations of the Lagrangian mean and Eulerian mean velocities $\ub^L$ and $\ub^E$, respectively, is a recurring issue in both the Craik-Leibovich (CL) and generalised Lagrangian mean (GLM) literatures. However, in the SCL analogue here between the It\^o correction velocity $\bu^S(\bx)$ (\emph{without} overline) and the Stokes mean drift velocity $\ub^S(\bx)$ (\emph{with} overline), we \emph{already know} the divergence of the It\^o correction $\bu^S(\bx)$ in equation \eqref{StokesDrift-ansatz} when we need to impose ${\rm div}( \bu^L_t)=0$ to determine the pressure semimartingale $\diff p$; in remark \ref{pressure-martingale}. }
See Remark \ref{pressure-martingale} below for the semimartingale formula which determines the pressure. To finish the notation, $g$ in the Lagrangian \eqref{lag-stoch} denotes the gravitational constant. 

Hamilton's principle with the stochastic constraints \eqref{aux-stoch} now yields a stochastic Kelvin-Newton theorem \cite{HMR1998}, expressible as, cf. \eqref{KelCirc-Strat},
\begin{align}
{\diff\,} \oint_{c(\diff \bx^L_t)} \!\!\! \bu_t \cdot d{\bf x}
= -\ g\oint_{c(\diff \bx^L_t)}  \!\!\! b\,dz\,dt
-\ \oint_{c(\diff \bx^L_t)}  \!\!\! d \Big(	{\rm d}p - \frac12 |\bu_t|^2\,dt + \frac12|\bu^S(\bx)|^2\,dt\Big)
\,,
\label{Kel-stoch-u}
\end{align}

in which $\bu_t:=\bu^L_t-\bu^S(\bx) $ and the closed loop $c(\diff \bx^L_t)$ moves with velocity $\diff \bx^L_t$ of the Lagrangian trajectory in \eqref{strat-traj}. Again, the last term will vanish in the Kelvin-Newton theorem \eqref{Kel-stoch-u}. 

When $\bu^S$ vanishes, equation \eqref{Kel-stoch-u} yields Kelvin's circulation theorem for the stochastic Euler-Boussinesq (SEB) equations. Remarkably, though, when $\bu^S$ is finite, as given in \eqref{StokesDrift-ansatz}, equation \eqref{Kel-stoch-u} yields Kelvin's circulation theorem for the stochastic Craik-Leibovich (SCL) equations, whose deterministic version (DCL) is used for modelling Langmuir circulations in the oceanic mix layer \cite{C85,CL1976}. 

Being loop integrals of exact differentials, the last terms in equations \eqref{KelCirc-Strat} and \eqref{Kel-stoch-u} both vanish. However, including the last term allows us to envision the SCL equations in full. Namely, for the Lagrangian trajectory $\diff \bx^L_t$ in equation \eqref{strat-traj}, applying the KIW formula \eqref{KelCirc-Strat} to the Kelvin circulation integral on the left side of equation \eqref{Kel-stoch-u} yields the stochastic motion equation, as
\begin{align}
{\diff\,} \bu_t  - \diff \bx^L_t \times {\rm curl} \bu_t 
+ \nabla \big(\diff \bx^L_t \cdot \bu_t \big)
= -\ g b\,\nabla z\,dt
- \nabla  {\rm d}p - \nabla \Big(- \frac12 |\bu_t|^2 + \frac12|\bu^S(\bx)|^2\Big)dt
\,.
\label{SCL-eqns-u}
\end{align}
The SCL motion equation \eqref{SCL-eqns-u}  includes all three of the velocities $\bu_t$, $\bu^L_t$ and $\bu^S_t$. Although the velocities are mixed in this equation, it implies a compact version of the Kelvin circulation theorem,
\begin{align}
{\diff\,} \oint_{c(\diff \bx^L_t)} \!\!\! \bu_t \cdot d{\bf x}
= -\ g\oint_{c(\diff \bx^L_t)}  \!\!\! b\,dz\,dt
\,,
\label{Kel-stoch-Eulvel-u}
\end{align}
where the closed loop $c(\diff \bx^L_t)$ is transported by the stochastic vector field $\diff \bx^L_t$ in \eqref{strat-traj} and the integral of gradients around the closed loop have vanished. 
As we have discussed, in the physical understanding of the Kelvin circulation theorem, one should regard the velocity $\bu_t$ in the integrand as an Eulerian quantity and the flow velocity $\diff \bx^L_t$ of the material loop  as a Lagrangian quantity.

\begin{remark}[Determining the pressure semimartingale]\label{pressure-martingale}
To determine the pressure semimartingale $({\rm d}p)$ one imposes preservation of ${\rm div}\bu_t=0$ on the divergence of the  motion equation \eqref{SCL-eqns-u} to find a semimartingale Poisson equation for ${\rm d}p$,
\begin{align}
\Delta \bigg({\rm d}p + \diff \bx^L_t \cdot \bu_t  + \Big( - \frac12 |\bu_t|^2 + \frac12|\bu^S(\bx)|^2\Big)dt\bigg)
=
 {\rm div} \Big(  \diff \bx^L_t \times {\rm curl} \bu_t  -\ g b\,\nabla z\,dt  \Big)
\,,
\label{dp-eqn}
\end{align}
with Neumann boundary conditions obtained by preservation of the condition that $\bu_t$ have no normal component on the fixed boundary of the flow domain. For an explanation of why the pressure must be regarded as a semimartingale for ${\rm \textcolor{red}d}p$ to impose incompressibility on a stochastic vector field, see \cite{SC2020}.  \hfill $\Box$
\end{remark}

\begin{remark}[Completing the stochastic dynamical system]
The SCL motion equation \eqref{SCL-eqns-u} is completed by the auxiliary stochastic advection equations for $b$ and $D$ in equation \eqref{aux-stoch}. The constraint $D-1=0$ imposed by the Lagrange multiplier ${\rm d}p$ (the pressure semimartingale) in \eqref{lag-stoch} ensures that the velocity $\bu^L_t $ is divergence free, provided the drift velocity $\bu^S(\bx)$ in \eqref{StokesDrift-ansatz} also has no divergence.  \hfill $\Box$
\end{remark}

Equation \eqref{SCL-eqns-u} may be equivalently written in terms of only $\bu^L_t$ and $\bu^S_t$ as
\begin{align}
\begin{split}
{\diff\,} \bu^L_t - \diff \bx^L_t \times {\rm curl} \bu^L_t 
+ \nabla \big(\diff \bx^L_t \cdot \bu^L_t  \big)
= &-\ g b\,\nabla z\,dt
 + \diff \bx^L_t \times {\rm curl} \bu^S(\bx) 
\\&- \nabla \Big( {\rm d}p + \diff \bx^L_t \cdot \bu^S(\bx)\Big) + \nabla \Big( \frac12 |\bu^L_t - \bu^S(\bx)|^2 - \frac12|\bu^S(\bx)|^2
\Big)dt
\,,
\end{split}
\label{SCL-eqns}
\end{align}
where we have dropped the term ${\diff\,} \bu^S(\bx)$ because $\bu^S(\bx)$ in equation \eqref{StokesDrift-ansatz}  is time-independent. The remaining terms involving $\bu^S(\bx)$ comprise a stochastic version of the `vortex force' in DCL and an added stochastic contribution to the pressure. This vortex force appears in the corresponding Kelvin theorem as a source of circulation of the velocity $\bu_t^L$, viz.,
\begin{align}
{\diff\,} \oint_{c(\diff \bx^L_t)} \!\!\! (\bu_t^L - \bu^S(\bx)) \cdot d{\bf x}
= -\ g\oint_{c(\diff \bx^L_t)}  \!\!\! b\,dz\,dt 
\,.
\label{Kel-stoch-Eulvel}
\end{align}
The ``vortex force" of the Deterministic Craik-Leibovich (DCL) theory was introduced to model the observed phenomenon of Langmuir circulations arising physically from wave--current interaction (WCI), \cite{L1}-\cite{LT}. The importance of including $\ub^S$ in the DCL equations is investigated for Kelvin-Helmholtz instability in \cite{Holm1996} and for symmetric and geostrophic instabilities in the wave-forced ocean mixed layer in \cite{HaneyF-K2015}. The results of having made the ``vortex force" of the SCL theory stochastic have yet to be investigated in solutiopns of the 3D SEB equations. 

Equation \eqref{SCL-eqns} with $\bu_t:=\bu^L_t-\bu^S $ is an example of our earlier discussion after equations \eqref{Kel-forcelaw-conc1} and \eqref{Kel-accelaw-conc1} in which the acceleration figures in the Kelvin-Newton relation, because the specific momentum $\bm{u}_t(\bx)$ is linear in the fluid transport velocity $\bm{u}_t^L(\bx)$ at fixed points in Eulerian coordinates and with time-independent coefficients. In this case, equations \eqref{SCL-eqns} and \eqref{Kel-stoch-Eulvel} exemplify the $a=F/m$ version of Newton's law which arises in this special case. Thus, the stochastic ``vortex force" in equation \eqref{Kel-stoch-Eulvel} is a \emph{non-inertial force} which arises from insisting on writing the acceleration of the relative velocity instead of the rate of change of momentum in Newton's force law. The stochastic motion equation \eqref{SCL-eqns-u} has no non-inertial ``vortex force", because it is written entirely in the Eulerian data frame. The non-inertial ``vortex force" arises in equation \eqref{SCL-eqns} upon replacing rate of change of Eulerian specific momentum $\bu_t$ in \eqref{SCL-eqns-u} with rate of change of the Lagrangian transport velocity (Lagrangian acceleration) $\bu^L_t$ in equation  \eqref{SCL-eqns}. 

\subsection{Vorticity and PV dynamics}
The curl of the SCL motion equation \eqref{SCL-eqns} yields the dynamics for the \emph{total vorticity} 
\begin{align}
\bm{\omega}_t :={\rm curl}(\bu^L_t - \bu^S ) = {\rm curl\,} \bu_t
\,,\label{SCL-total-vort-def}
\end{align}
which is given by 
\begin{align}
{\diff\,}\bm{\omega}_t - {\rm curl}\big(\diff \bx^L_t \times \bm{\omega}_t \big)
= - g \nabla b \times \nabla z
\,.
\label{SCL-total-vort}
\end{align}
The total vorticity dynamics \eqref{SCL-total-vort}, in turn, yields a stochastic advection law for the \emph{total potential vorticity}, defined by $q:=\bm{\omega}_t\cdot\nabla b$; namely, 
\begin{align}
 {\diff\,} q + \diff \bx^L_t \cdot \nabla q = 0
\,.\label{PV-stoch-advec}
\end{align}
In turn, this  implies preservation of spatial integrals 
\begin{align}
C_\Phi = \int_{\cal D} \Phi (q,b) \,d^3x
\,,\label{PV-Casimirs}
\end{align}
for arbitrary differentiable functions $\Phi$, provided $\diff \bx^L_t$ has no normal component at the boundary $\p {\cal D}$ of the flow domain ${\cal D}$. 

\section{Conclusion} 

The central theorem for fluid dynamics (the Kelvin theorem) involves two frames in which velocities are measured. As we have seen, one velocity is a vector and the other is a co-vector. The integrand is in a fixed inertial  frame and the circulation loop is in the moving frame of the Lagrangian fluid parcels. The frame of the specific momentum in the integrand is Eulerian and the frame of the moving loop is Lagrangian. Likewise the data observation frame and the fluid motion frame will differ, if one is modelled as It\^o and the other as Stratonovich. Thus, it makes sense that the shifts between frames  which occur in transforming a Lagrangian trajectory from It\^o to Stratonovich form would introduce non-inertial forces in the motion equations. This was already clear from the Coriolis force and the Craik-Leibovich vortex force in the deterministic modelling of fluid dynamics. 

Similarly, waves are Eulerian while fluid motion is Lagrangian: waves move relative to fixed space through the moving fluid, while the motion of the fluid Doppler shifts the wave frequency. In the Craik-Leibovich model, the Eulerian velocity (defined as the total specific momentum) is given by $\ub_t=\ub_t^L-\ub^S(\bx)$.  This is the difference between the Lagrangian fluid transport velocity $\ub_t^L$ and another velocity $\ub^S(\bx)$ called the Stokes drift velocity due to the waves, which must be prescribed from observed wave conditions. The Craik-Leibovich non-inertial vortex force arises as in \eqref{SCL-eqns} for the same reason as the Coriolis force arises in equations \eqref{Kel-forcelaw-conc1} and \eqref{Kel-accelaw-conc1}, except that one replaces $\bm{R}(\bx)\to-\,\ub^S(\bx)$. Namely, the acceleration (i.e., the time rate of change of the circulation of $\bms{u}_t^L(\bx)$ the fluid velocity relative to the moving frame) equals the sum of the force in the inertial frame $\bm{F}$, plus the non-inertial force $\bm{F}_{Coriolis}$. 

\subsection*{Conclusion: does the It\^o correction make a difference? \\ 
Answer: No, for total specific momentum, Yes, for relative velocity. }

What does all this mean for the original problem of comparing It\^o data  with Stratonovich equations of motion derived from Hamilton's principle for stochastic fluid equations in the Euler-Poincar\'e form \eqref{EP-stoch}? It means that no non-inertial forces due to changes of frame by the It\^o correction need to be considered for the dynamics of the total specific momentum, $\bm{u}_t=\bm{u}_t^L - \frac12 \big(\bm{\xi}(\bx_t)\cdot\nabla\big) \bm{\xi}(\bx_t)$, which lives naturally in the Eulerian \emph{data frame}. However, if one seeks the dynamics of the Lagrangian relative transport velocity, $\bm{u}_t^L$, instead of the Eulerian specific momentum, $\bm{u}_t$, then non-inertial forces will arise due to the It\^o-Stokes correction, $\ub^S(\bx_t)=-\frac12 \big(\bm{\xi}(\bx_t)\cdot\nabla\big) \bm{\xi}(\bx_t)$. 

\subsection*{Acknowledgements.} I am enormously grateful for many illuminating discussions about stochastic data assimilation with B. Chapron, C. J. Cotter, D. Crisan, E. Mem\'in, W. Pan and  I. Shevchenko over the past few years. Thanks also to E. Luesink for coining the term ``data frame''  and to O. Street for raising the issue of interpreting the pressure in the SEB/SCL equation \eqref{SCL-eqns-u} as a semimartingale. This work was partially supported by EPSRC Standard grant EP/N023781/1 and by ERC Synergy Grant 856408 - STUOD (Stochastic Transport in Upper Ocean Dynamics).

\end{document}